# High Rate CV-QKD Secured Mobile WDM Fronthaul for Dense 5G Radio Networks

**Dinka Milovančev, Nemanja Vokić, Fabian Laudenbach, Christoph Pacher, Hannes Hübel, and Bernhard Schrenk**

*AIT Austrian Institute of Technology, Center for Digital Safety&Security / Security & Communication Technologies, 1210 Vienna, Austria.*
  Author e-mail address: bernhard.schrenk@ait.ac.at

A coherent transmission methodology for a continuous-variable quantum key distribution (CV-QKD) system based on quantum-heterodyne measurement through a coherent intradyne receiver is experimentally demonstrated in the framework of 5G mobile fronthaul links. Continuous optical carrier synchronization is obtained through training information, which is multiplexing to the quantum signal as pilot tone in both, frequency and polarization. Spectral tailoring by means of optical carrier suppression and single-sideband modulation is adopted to simultaneously mitigate crosstalk into the quantum channel and self-interference for the pilot tone, thus allowing for a high signal-to-noise ratio for this training signal. Frequency offset correction and optical phase estimation for the free-running local oscillator of the receiver is accurately performed and guarantees low-noise quantum signal reception at high symbol rates of 250 MHz and 500 MHz with additional Nyquist pulse shaping. A low excess noise in the order of 0.1% to 0.5% of shot-noise units is obtained for fiber-based transmission over a fronthaul link reach of 13.2 km. Moreover, co-existence with 11 carrier-grade classical signals is experimentally investigated. Joint signal transmission in the C-band of both, quantum signal and classical signals, is successfully demonstrated. Secure-key rates of 18 and 10 Mb/s are obtained under strict security assumptions, where Eve has control of the receiver noise, for a dark and a lit fiber link, respectively. Moreover, rates of 85 and 72 Mb/s are resulting for a trusted receiver scenario. These secure-key rates are well addressing the requirements for time-shared CV-QKD system in densified 5G radio access networks with cloud-based processing.

## I. Introduction

The fifth generation of wireless communications (5G) has introduced disruptive performance changes to mobile networks [1]. Enhanced mobile broadband promises peak data rates of up to 10 Gb/s in virtue of elevated carrier frequencies in the mm-wave band and beamforming. Low latency and ultra-high reliability ensure the deployment of mission-critical services over wireless networks. Lastly, the realization of a highly densified radio access network supports a massive user count, as required to unleash small cell paradigms and the Internet-of-Things. These performance characteristics render 5G radio as a key asset in the global telecommunication infrastructure. However, 5G can hardly be deployed without optical layer. Radio signals that are received in an interference-limited regime are calling for extensive signal processing. Cloud architectures support the centralized and cooperative processing of radio signals from remote antenna sites, for which the waveforms are optically hauled to the processing warehouses [2]. This optical backbone between a centralized point of presence and the antenna sites should enjoy highest security standards in terms of data and control.

Quantum physics offers the required methods to ensure information-theoretic secure data encryption and protection of the communication infrastructure, in the form of quantum key distribution (QKD). One of the main technological challenges remaining is the single-photon detection mechanism. In practical discrete-variable (DV) implementations that are based on single-photon encoding, the reception of optical quantum signals builds on very specific detectors based on either avalanche photodiodes operated in Geiger-mode or super-conducting nano-wires [3, 4]. However, these detectors are either slow or come at a degree of complexity that renders them as cost-ineffective to integrate with telecom-grade systems.

Continuous-variable (CV) quantum communication is considered as a contender with credentials to tackle this preclusion. It evades the need for Geiger-mode photodetection by means of coherent optical reception of highly attenuated coherent states. CV-based QKD, as a representative application of quantum communication technology, exploits the Heisenberg uncertainty of coherent quantum states to securely generate random keys among two end-nodes [5]. The use of coherent reception fits squarely in the roadmap of system integrators. The technological overlap with optoelectronics as found in telecom and datacom transceivers enables access to a higher bandwidth, which is essential in view of growing transmission capacities.

In this work we experimentally demonstrate a coherent transmission methodology for a 250-Mbaud coherent intradyne CV-QKD system with frequency and polarization-multiplexed training signal for local oscillator (LO) synchronization. A pilot tone is spectrally tailored for this purpose. Minimal crosstalk to the quantum signal is ensured through rigorous single-sideband modulation and optical carrier suppression. Co-existence with 11 carrier-grade classical channels is validated for joint C-band transmission of quantum and classical signals over a mobile fronthaul link with a reach of 13.2 km. Furthermore, we investigate the use of Nyquist pulse shaping to double the CV-QKD symbol rate to 500 Mbaud to best exploit the available receiver bandwidth in view of possible time-sharing of QKD equipment. High secure-key rates of 85.3 and 72 Mb/s result for the dark and lit fiber link, respectively.

The paper is organized as follows. Section II introduces the QKD-secured 5G architecture and its requirements in terms of secure-key rate. Section III relates the CV-QKD system to the state-of-the-art and elaborates on the synchronization of the free-running LO. It further discusses the suppression of crosstalk due to the addition of synchronization information. Section IV details the experimental setting and the signal characteristics. Section V discusses the results, the performance impact of pulse shaping and the co-transmission of classical channels. Section VI finally concludes this work.

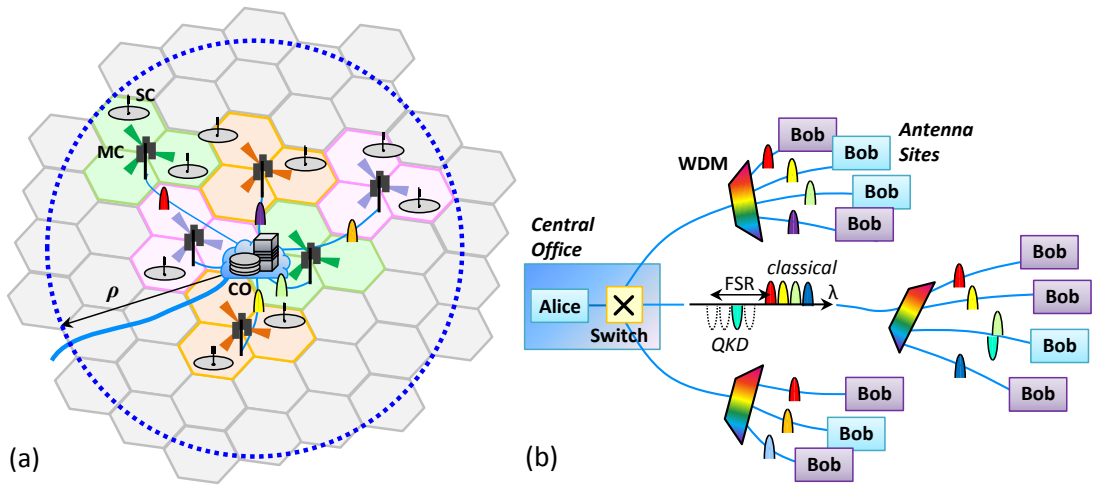

Fig. 1. (a) 5G radio access architecture. (b) QKD-secured mobile fronthaul.

## II. QKD-Enabled Security in 5G Optical Networks

### A. A Quantum-key secure mobile optical fronthaul

The communication network enables pervasive information exchange between industrial, business and residential end-users as well as between machines. It is considered as a critical infrastructure and should therefore adhere to the highest security standards. Under this premise, securing the control plane between field-distributed and centralized radio access network assets through means of QKD is under intense discussion. It is also the prime focus of this work, which on top questions whether the data traffic can be secured at the same time by means of high accomplished QKD key rates.

The 5G infrastructure follows a cloud-based radio access network architecture [2], which is introduced in Fig. 1(a). The signals from the remote radio heads (RRH) at the antenna sites are optically fronthauled to centralized signal processors, located at the central office (CO) or dedicated mobile edge computing sites. In this way signal processing can be efficiently pooled at dedicated datacenter hostels and cooperative multipoint schemes can be adopted. The link reach $\rho$ to the antenna sites is strictly limited to less than 20 km through the demanding 5G latency specs. Moreover, an optimized mix of macro-cells (MC) and small cells (SC) is employed to ensure a densified radio network.

Mobile fronthauling can be deployed according to multiple functional split options [6], which determine the degree of signal processing that is performed at the RRH. In the extreme and most demanding case, that the native radio signal is digitized and forwarded to the CO for further processing, the highest capacity is required for the mobile optical fronthaul and can peak at 20 Gb/s per beam and GHz of radio bandwidth [7]. To support such a fronthaul, a wavelength division multiplexed (WDM) overlay has been specified in modern access standards such as NG-PON2 [8], in order to provide virtual point-to-point links between CO and RRH at the optical distribution network.

| Network resource | Scenario / Characteristic | Macro-Cell | Small Cell |
|---|---|---|---|
| RRH | Data rate per beam | 20 Gb/s | 100 Gb/s |
| | Multi-user MIMO | 5 beams | 1 beam |
| Antenna Site | Sectors | 3 | 1 |
| | Inter-site distance | 200 m | - |
| | Small cells per macro-cell sector | 4 | - |
| Mobile Fronthaul | Reach $\rho$ | 20 km | |
| | Spectral allocation | C-band | |
| QKD | AES key update rate | 1 / 64 Gbyte | |
| | Spectral allocation | C-band | |

Table I. Radio access network parameters.

End-user data and control signals are jointly transmitted over the mobile fronthaul. A fully quantum-secured fronthaul can be accomplished in co-existence with this classical information, provided that the adopted QKD methodology is robust to Raman-induced noise by the classical WDM channels and that the accomplished secure-key rate is high enough to support the fronthaul capacity deriving from the 5G radio signals. Architecture wise, the WDM-based passive optical network (PON), sketched in Fig. 1(b), can be advantageously re-used for the quantum layer. CV-QKD can be established at a parallel virtual point-to-point link, for example by exploiting the free spectral range (FSR) property of WDM multiplexers. As we will prove in this work, CV-QKD transmission is robust even when choosing a QKD wavelength in the C-band, and thus in the same waveband as the classical channels. In practice, the densified 5G network and the high number of antenna sites within the reach $\rho$ will lead to many WDM PONs being connected to a single CO. In case that the secure-key rate is high enough, a few or even a single CV-QKD head-end at the CO can be shared among the antenna sites by time-sharing the CO unit. This is illustrated in Fig. 1(b), where one head-end Alice is connected to many tail-end Bobs by means of spatial switching and spectral multiplexing. By selecting the feeder fiber through the spatial switch at the CO and by tuning the QKD wavelength within the WDM grid, every Bob can be sequentially addressed. Setting and breaking up point-to-point QKD links between Alice and any Bob can be made on a circuit rather than on a packet level. It does not require fast switching since an established secure key can be buffered at both link ends.

It shall also be stressed that Alice and Bob can be interchangeably placed at CO and antenna site and that classical communication associated to the authorization and sifting process can be integrated in the auxiliary management and control of the classical data channel. Moreover, a migration of optical access and mobile fronthauling to coherent PON technology supports the integration of CV transceiver functions in virtue of multi-purpose photonics. This allows to smoothly introduce QKD without resorting to specialized single-photon detectors.

It is clear that, unless an optical wireless link is implemented, a QKD pipe is not available between the RRH and the mobile end-user. In order to extend quantum-safe encryption to the wireless RF domain, one has to resort to post-quantum encryption towards a hybrid QKD/post-quantum end-to-end implementation between mobile user and CO [9].

*B. Required secure-key rate*

An estimation on the required key rate has been conducted based on the parameters listed in Table I [10-12]. The overhead due to network control is considered as negligible when compared to the traffic arising from the end-user data. The CPRI-equivalent traffic generated for a single antenna beam amounts to 20 Gb/s due to oversampling and resolution for a radio signal bandwidth of 1 GHz [7]. Moreover, we consider that a MIMO antenna configuration allows multi-user beamforming to serve up to 5 users with independent data streams. With this, a single antenna site is associated to a CPRI-equivalent data rate of 300 Gb/s for a tri-sectoral macro-cell. Omnidirectional small cells are dedicated to a single user, however, they may exploit mm-wave frequencies with a bandwidth of 5 GHz. This relates to a data rate of 100 Gb/s. The inter-site distance for macro-cells shall be 200 m, whereas four small cells may exist per macro-cell sector. In a fully densified radio network this corresponds to a total of 4000 macro-cells and, accordingly, 48k small cells within the latency-limited coverage of a CO.

Clearly, the established QKD key cannot support one-time pad encryption since the secure-key rate is much lower than the classical data rate. However, it can be used to periodically change the AES key in a symmetric encryption setting. In such a scheme, a low attack success probability of ~$2^{-50}$ can be expected if the AES key is updated after transmission of a 64 Gbyte data chunk [13]. Given the introduced mobile fronthaul architecture, the QKD system is required to deliver an aggregated secure-key rate of 3 Mb/s in order to ensure this periodic AES key change for all antenna stations.

Furthermore, it has to be recalled that a time-shared QKD system requires synchronization whenever a transmitter-receiver pair is switched. Although this reduces the duty cycle in which a key is established further, an elevated key rate for the CV-QKD system and a sufficiently large key storage can account for this downtime. Classical coherent communications can build on fast tunable lasers [14] and fast carrier locking [15] for this purpose. Let us assume a synchronization time of 5 s due to wavelength re-tuning and transmitter-receiver synchronization for the QKD system. The nominal secure-key rate shall be elevated to 10 Mb/s and the key storage has a size of 200 Mbit. With the peak data rate of 300 Gb/s for an antenna site as discussed previously and the rates for generation and consumption of key, it follows that the filled key buffer is eroded by 97.5% at the time when the storage of the antenna is again fed with freshly generated key. This means that extensive key buffering and high secure-key rates can both account for the switching inefficiency in trade-off with the simplicity of deploying just a single head-end QKD system. While buffering can be applied for any QKD system architecture, high key rates can be supported through CV-QKD, which is the focus of this work.

| Ref. | Year | true LO | LO Scheme / Training | Reception | Link reach [km] | TX rate | QKD λ [nm] | Classical λ [nm] | DSP | ξ [SNU] | SKR |
|---|---|---|---|---|---|---|---|---|---|---|---|
| [22] | 2012 | ✗ | TI+PDM LO | homodyne | 1.7, SMF | 500 kHz | 1550 | 1310, 1490 | real-time | 0.05 | 600 b/s |
| [23] | 2013 | ✗ | TI+PDM LO | homodyne | 80, SMF | 1 MHz | 1550 | - | offline | 0.002 | 100 b/s |
| [34] | 2015 | ✓ | TI ref. pulses | homodyne | 25, SMF | 100 MHz | 1550 | - | offline | 0.05 | 100 kb/s |
| [24] | 2015 | ✗ | TI+PDM LO | homodyne | 25, SMF | 50 MHz | 1550 | 1570/90,1610 | offline | 0.10 | 1 Mb/s |
| [25] | 2015 | ✗ | TI+PDM LO | homodyne | 75, SMF | 1 MHz | 1531.12 | 1550.12 | offline | 0.0016 | 0.5 kb/s |
| [26] | 2016 | ✗ | TI+PDM LO | homodyne | 100, SMF | 2 MHz | 1550 | - | real-time | 0.015 | 500 b/s |
| [27] | 2016 | ✗ | TI+PDM LO | homodyne | 17.5, SMF | 500 kHz | 1550.12 | 1570/90,1610 | offline | N/A | 0.25 kb/s |
| [28] | 2017 | ✗ | TI+PDM LO | homodyne | 10, SMF | 10 MHz | 1550 | - | real-time | 0.02 | 50 kb/s |
| [35] | 2017 | ✓ | 2 FDM pilots | homodyne | 40, SMF | 40 MHz | 1558 | - | offline | 0.016 | 240 kb/s |
| [29] | 2018 | ✗ | TI+PDM LO | homodyne | 10, SMF | 10 MHz | 1549.6 | 18λ, 1544-48 | real-time | N/A | 75 kb/s |
| [33] | 2018 | ✓ | TI ref. pulses | homodyne | 25, SMF | 50 MHz | 1542.38 | - | offline | 0.04 | 3.1 Mb/s |
| [37] | 2018 | ✓ | FDM pilot | heterodyne | 20, SMF | 10 MHz | 1550 | 10λ, C-band | offline | N/A | 90 kb/s |
| [38] | 2019 | ✓ | FDM/PDM pilot | heterodyne | 40, SMF | 250 MHz | 1550.12 | - | offline | 0.04 | 3 Mb/s |
| [39] | 2019 | ✓ | 2 FDM pilots | heterodyne | 25, ULL | 40 MHz | 1504.98 | 56λ C-band + weak 1509 | offline | 0.0037 | 400 kb/s |
| [30] | 2019 | ✗ | TI+PDM LO | homodyne | 10, SMF | 10 MHz | 1550 | 100λ, C-band | offline | N/A | 29 kb/s |
| [40] | 2020 | ✓ | FDM/PDM pilot | heterodyne | 25, SMF | 500 MHz | 1527 - 1566 (194 ch.) | - | offline | 0.06 | 0.89 Mb/s (per ch.) |
| [41] | 2020 | ✗ | self-homodyne | heterodyne | 15, SMF | 250 MHz | 1549.73 | - | offline | 0.08 | 14.2 Mb/s |
| [42] | 2020 | ✓ | FDM/PDM pilot | heterodyne | 13.2, SMF | 500 MHz | 1532.9 | 11λ, 1548-60 | offline | 0.0033 | 22.3 Mb/s |
| [43] | 2020 | ✗ | self-homodyne | heterodyne | 13, ULL | 25 MHz | 1550 | 1542 | offline | 0.03 | 0.3 Mb/s |

Quantum homodyne reception corresponds to coherent intradyne reception with π/2 hybrid. Quantum heterodyne reception corresponds to either coherent intradyne reception in baseband using a π/4 hybrid, or to coherent heterodyne reception using a π/2 hybrid.

Abbreviations and symbols used in the table: TI time-interleaved, TDM time division multiplexing, FDM frequency division multiplexing, ULL ultra-low loss fiber, ξ excess noise, SKR secure-key rate.

Table II. Experimental CV-QKD system demonstrations

## III. CV-QKD with Carrier-Suppressed Pilot Tone

*A. The State-of-the-art in CV-QKD*

CV-QKD not only benefits from mature opto-electronics of coherent telecommunication systems. Coherent detection also contributes to narrowband opto-electronic filtering, which makes CV-QKD systems more robust to Raman scattering when classical channels co-exist on the same transmission span [16]. DV-based quantum systems reject Raman noise through the adoption of optical filters as they are applied in wavelength division multiplexing (WDM) architectures [17] such as they apply in 5G networks, whereas coherent detection enjoys a much steeper opto-electronic filtering response. For example, direct photodetection in the remote O-band is restricted to commercial standards such as coarse WDM or local area network (LAN) WDM, leading to reception bandwidths of

~500 GHz. Early experimental co-existence studies have revealed that these filters pass too much noise to the single-photon receiver in case of carrier-grade classical channels [18, 19] and had resorted to highly attenuated classical channels for QKD co-transmission, unless custom filters had been adopted more recently [20]. On the contrary, coherent detection reduces the impact of Raman noise to the actual opto-electronic receiver bandwidth in the sub-GHz regime, thus gaining robustness for a joint transmission of closely spaced quantum and classical channels.

Nevertheless, the practical realization of CV-QKD requires coherent receivers to operate at <1 photon/symbol, at an excess noise of ~1% of shot-noise units of its LO, and is therefore linked to certain technological challenges. Early implementations of CV-QKD systems relied on the joint transmission of the quantum signal and the LO. In such a scheme [21-30], a pulsed LO is time-interleaved with the quantum symbols and demultiplexed at the receiver for the purpose of synchronized coherent detection. Since the LO power at the receiver is dependent on the channel parameters, such an implementation proves difficult if increased channel loss is to be overcome. Moreover, the co-transmitted LO can be subject to eavesdropping attacks.

Concepts employing a receiver-side LO [31-44], as it is commonly seen in telecom-based coherent transceivers, have gained significant attention during recent years. Challenges such as the optical carrier synchronization of now independent optical sources at transmitter and receiver have been solved through incorporation of auxiliary training information and powerful signal processing resources.

A variety of methods to train such a CV receiver towards optimal operation have been proposed. These include to discipline its free-running LO through crosstalk-free multiplexing of synchronization information, rather than building on classical data-aided signal recovery schemes. Demonstrations of such schemes have adopted time-interleaving [31-34] for its training signals in combination with additional polarization multiplexing, frequency multiplexed [35-37, 39] or frequency/polarization multiplexed pilots [38, 40, 42] or tailoring of the modulation alphabet [44]. Table II compares the obtained results for various CV-QKD systems. The highest transmitted symbol rates are obtained for CV-QKD systems employing continuous training information multiplexed in frequency and quantum heterodyne detection in the baseband, coherent intradyne reception [38, 40-42].

Towards this direction, we have recently proposed an LO synchronization scheme that employs a continuous pilot tone multiplexed in polarization and frequency and have validated this system over a dark fiber link [38]. We build on a quantum heterodyne measurement of I/Q quadratures through coherent intradyne reception. Compared to coherent heterodyne reception in a passband that shifted from the low-frequency region, as in [37] or [39], this reception methodology allows us to exploit the precious quantum receiver bandwidth more effectively, thus allowing higher transmission rates. This work extends our very recent related CV-QKD investigation, which incorporates quantum symbol shaping over a lit channel [42], in order to enable a secure-key rate as high as possible. In the present work, we give further insight on the concept behind the CV-QKD system, relate the accomplished QKD key rates to 5G fronthaul applications and investigate the importance of selected system parameters, such as the reception bandwidth and the detector balancing in presence of co-existing classical channels.

*B. CV-QKD with carrier-suppressed, single-sideband pilot*

The principal concept behind the CV-QKD system that is studied in this work is presented in Fig. 2 and follows the methodology of a pilot tone that is multiplexed in the frequency and polarization dimensions. There is no QKD protocol presented in this work but rather a transceiver architecture that can be applied to different CV-QKD protocols. In the present case, the CV system builds on a four-state quantum protocol implemented in the optical phase space. This protocol has been chosen for the sake of simplicity, though a deployable CV-QKD system would be based on Gaussian modulation. Nevertheless, the presented synchronization and modulation is transparent to the protocol.

A quantum-heterodyne measurement at the receiver is implemented through coherent intradyne detection, which directly yields the in-phase (I) and quadrature (Q) variables. These variables are corrected in their frequency offset and optical phase drift between the sources at transmitter and receiver by means of digital signal processing (DSP). To obtain the quantum data at low excess noise $\xi$ after signal processing, auxiliary training information is communicated through an out-of-band pilot tone, which is polarization division multiplexed (PDM) to the quantum signal.

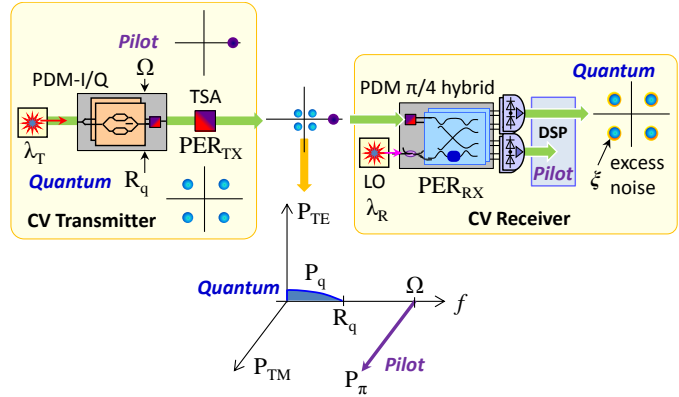

Fig. 2. Pilot-disciplined CV-QKD system building on training information multiplexed in frequency and polarization.

Figure 3 explains the methodology for CV transmission based on quantum data and pilot tone. The baseband four-state quantum signal (A) in TE polarization is realized through quadrature phase shift keying at a symbol rate of $R_q$ = 250 MHz < $B_{RX}$, where $B_{RX}$ is the opto-electronic bandwidth of the quantum receiver. The quantum receiver bandwidth is limited due to the specific low-noise design, as we will shortly show. Additional pulse shaping at the electrical modulator feed can be applied to further boost the symbol rate [46], as it will be investigated in this work for the particular case of Nyquist pulse shaping at $R_q$ = 500 MHz. In this way the quantum receiver bandwidth is exploited more efficiently and allows a higher transmit rate compared to previous works (Table II).

The optical launch power $P_q$ for the quantum signal has to satisfy the condition $S_q < P_q < P_{Eve} \ll P_{Sat}$, where $S_q$ and $P_{Sat}$ are the transmission-corrected sensitivity and saturation power of the quantum receiver, and $P_{Eve}$ the power level at which the quantum tributary becomes susceptible to eavesdropper attacks. This limits the launch $P_q$ to a typical value of 1...10 photons/symbol, depending on the exact channel length [45]. Further security features may address the removal of information from the transitions between the symbols by means of pulse carving at the rate $R_q$.

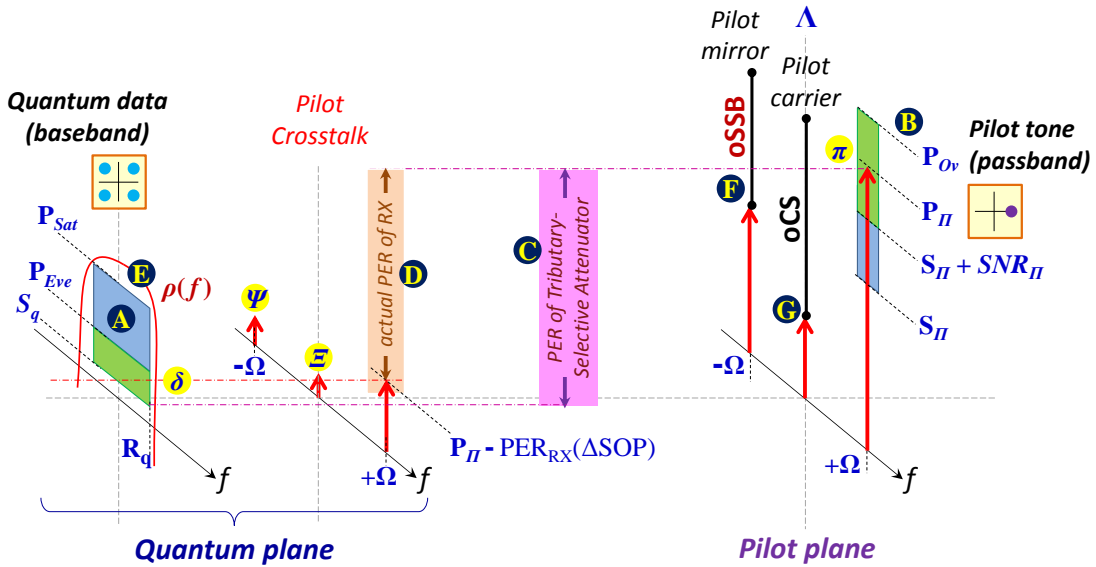

Fig. 3. Spectral allocation of quantum signal and pilot tone, respective power levels and crosstalk sources.

The pilot tone (B), which is multiplexed in TM polarization to the quantum signal for its recovery in frequency and phase, features a frequency of $\Omega$ = 1 GHz $\gg R_q$. With this it principally falls out-of-band to the quantum data and the launch power $P_\Pi$ can be chosen higher than for the quantum signal. In particular, the sensitivity $S_\Pi$ of the receiver in the pilot plane (i.e., the polarization plane dedicated to the pilot tone) and its overload $P_{Ov}$, both as seen from the transmitter-side, limit the launch range. In order to perform the frequency and phase estimation that is linked to the quantum signal recovery accurately, a high signal-to-noise ratio (SNR) is required, which yields the condition $S_\Pi + \text{SNR} < P_\Pi < P_{Ov}$. The launch level used in the later experiment was $P_\Pi$ = -46 to -49 dBm, depending on the actual modulation setting, and the relative power difference to the weaker quantum signal was realized through a polarization-selective attenuator (C), which affects both tributaries differently. The polarization extinction ratio $\text{PER}_{TX}$ for the quantum tributary can be chosen as function of the input state-of-polarization (SOP) of this attenuator. It yields the desired power ratio $P_\Pi/P_q$ of 20 dB for this experiment and ensures operation in the allowed ranges $\delta,\pi$ for the quantum signal and the pilot tone, resulting in a high SNR for the pilot and an acceptably low residual bleed-through of the pilot tone into the quantum receiver.

The alignment of the incident polarization state to the optical axis of the coherent PDM receiver (D) is paramount to guarantee crosstalk-free operation in the quantum and pilot plane. Since the maximum polarization extinction of the PDM receiver may be smaller than that at the transmitter, $\text{PER}_{RX} < \text{PER}_{TX}$, a residual pilot tone is received in the quantum plane (i.e., the polarization plane dedicated to the quantum signal) at a power $P_\Pi - \text{PER}_{RX}(\Delta \text{SOP})$. This power of the crosstalk note depends on the mismatch of the incident SOP from the optical axis and has to be suppressed to avoid excess noise during reception of the quantum signal. The large frequency difference between $R_q$ and $\Omega$, together with the rapidly falling frequency response function $\rho(f)$ of the quantum receiver (E) towards $\Omega$, guarantee a negligible error. At the same time, optical single-sideband (oSSB) modulation of the pilot tone eliminates the mirror pilot at $-\Omega$, which would lead to RF fading after coherent intradyne detection at the pilot plane. It also minimizes its crosstalk $\Psi$ to the quantum channel, without loss of information in the pilot plane.

While in case of the quantum signal the rigorous antipodal phase modulation leads to a carrier-less four-state signal in the TE plane, the optical carrier of the pilot tone resides. Despite frequency multiplexing of the pilot tone at the TM plane, its optical carrier leads to in-band noise $\Xi$ for the quantum signal in case of SOP misalignment. The suppression of this spectral component, which arises from the strong pilot tone, is only governed by $\text{PER}_{RX}$, which quickly vanishes already for a small $\Delta\text{SOP}$. Optical carrier suppression (oCS) is therefore adopted in combination with oSSB to ensure minimal excess noise for the received quantum signal or, from a different viewpoint, a highest possible SNR for the pilot tone. Both, oSSB and oCS, make the pilot tone a wavelength-shifted replica of the optical carrier $\Lambda$. This wavelength shift can be achieved with a single, well-known I/Q modulator. The optical field $E_{pilot}$ for the pilot is

$$E_{pilot} = \frac{E_{in}}{2} \left[ \begin{array}{l} \cos\left(\dfrac{\pi(v_{i,RF} + V_{i,B})}{2V_{\pi,i}}\right) + \\ \exp\left(j\dfrac{\pi V_{\varphi,B}}{2V_{\pi,\varphi}}\right)\cos\left(\dfrac{\pi(v_{q,RF} + V_{q,B})}{2V_{\pi,q}}\right) \end{array} \right] \quad (1)$$

where $E_{in}$ is the optical field of the light that seeds the PDM I/Q modulator at an optical power $P_{in}$, and $v_{i/q,RF}$ and $V_{i/q,B}$ are the RF drive signals and bias points of the I and Q branches, respectively. The half-wave voltages for these arms are given by $V_{\pi,i/q}$ and are further being considered identical. A phase shift between I and Q quadratures is introduced through the phase shifter of the I/Q modulator. The respective bias and half-wave voltage are given by $V_{\varphi,B}$ and $V_{\pi,\varphi}$, respectively. For a bias of $V_{i/q,B} = -V_{\pi,i/q}$ and $V_{\varphi,B} = V_{\pi,\varphi}$, the pilot becomes

$$E_{pilot} = \frac{E_{in}}{2} \frac{\pi}{2V_\pi} \left[ v_{i,RF} + v_{q,RF} \right] \quad (2)$$

A sinusoidal RF modulation at the pilot tone frequency $\Omega$ and an electrical phase of 90 degrees between I and Q leads to

$$E_{pilot} = \frac{\sqrt{P_{in}}}{2} \frac{\pi}{2V_\pi} \exp\left[j(\omega_\Lambda + \Omega)t\right] \quad (3)$$

This pilot resembles a spectral line that is shifted by $\Omega$ from the optical frequency $\omega_\Lambda$ of the modulator seed light. It does therefore not overlap with the signal spectrum at the quantum plane, provided that $\Omega > R_q$.

## IV. Experimental Setup and Signal Processing Stack

The experimental setup for evaluating the CV-QKD system in a WDM feeder is presented in Fig. 4, together with representative signal characteristics. It includes the CV transmitter/receiver pair with the respective digital signal processing (DSP) resources and the classical signal load.

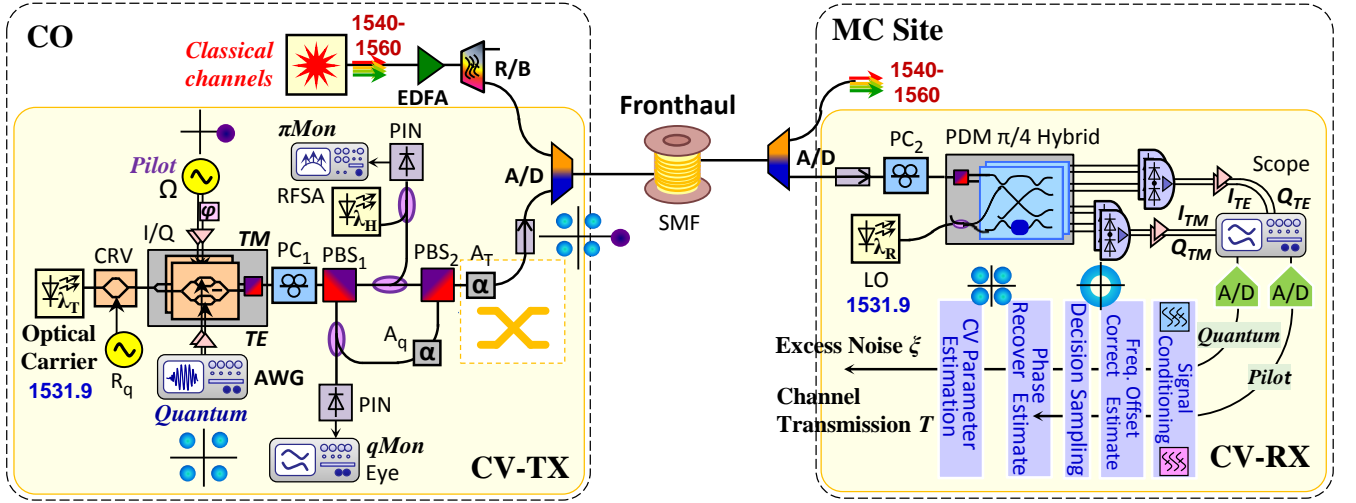

Fig. 4. Experimental setup for CV-QKD transmission with classical WDM overlay and DSP stack.

### C. Signal transmitter

The optical source at the CO-hosted CV transmitter delivers an optical carrier at $\lambda_T$ = 1531.9 nm with a linewidth of 10 kHz. The quantum signal and the pilot tone are jointly imprinted in both polarization tributaries, as described earlier. For this purpose a $LiNbO_3$ inphase/quadrature (I/Q) modulator in a polarization division multiplexed (PDM) configuration is used. This nested Mach-Zehnder I/Q modulator allows for four-state quantum signal generation and further accomplishes oCS-SSB modulation for the pilot tone, without loss of information integrity. Pseudo-random bit sequences with a length of $2^7-1$ have been used to feed the quantum signal generator. An additional pulse carver (CRV) is inserted between optical source and I/Q modulator in order to suppress the transition region between consecutive symbols of the quantum signal. The optical modulators were driven by RF synthesizers ($\Omega$, $R_q$) and an arbitrary waveform generator (AWG) in order to adopt pulse shaping for the data symbols so to investigate spectrally efficient methods for quantum signal transmission. In addition to the typically used Gaussian pulse filters, Nyquist pulse shaping is employed. It is based on a Raised-Cosine filter $H_{RC}$, for which the roll-off after the passband follows the function

$$H_{RC}(f) = \cos^2\left(\frac{2\pi f - (1-\alpha)\pi R_q}{4\alpha R_q}\right) \tag{4}$$

where $f$ is the frequency and $\alpha$ is the roll-off factor. Figure 5 presents a comparison between three pulse shaping filters $H_{RC}$. We have chosen a roll-off factor of $\alpha$ = 0.5 in view of the targeted symbol rate of 500 MHz and the available quantum receiver bandwidth, as discussed shortly.

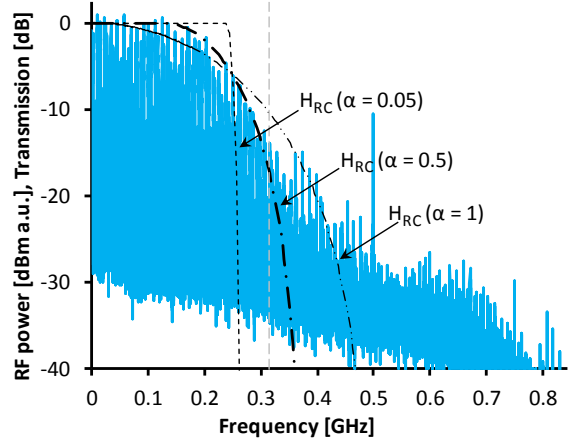

Fig. 5. Filter function for Nyquist pulse shaping for various roll-off factors α. The corresponding received quantum signal spectrum is shown for α = 0.5.

The respective eye diagrams for the transmitter-side quantum signal are included for uncarved (*α*) and carved (*β*) modulation in Fig. 6, together with those of the pilot tone (*γ*). These eye diagrams miss the null level at the symbol center, as it is characteristic for (carved) phase-modulated signals.

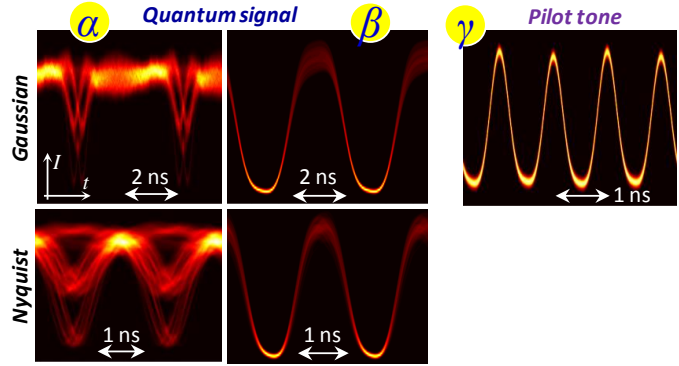

Fig. 6. Eye diagrams for the uncarved (α) and carved (β) quantum signal and the pilot tone (γ).

Signal monitors have been included for the quantum signal and the pilot tone and include an optically-preamplified data monitor (qMon) and an oCS-SSB monitor of the pilot tone (πMon) implemented through heterodyning with an additional optical reference ($\lambda_H$).

The power ratio between pilot tone and quantum signal is set through a tributary-selective attenuator. The attenuator arrangement splits the constituents through a polarization beam splitter (PBS) and attenuates ($A_q$) the quantum tributary before recombination to set the power ratio between pilot and quantum signal. Figure 7(a) shows the native electrical pilot tone spectrum at πMon after heterodyning with a reference laser. The optical carrier of the pilot signal is shifted to the heterodyne frequency $f_v$ = 5.77 GHz, which is given by the detuning of the reference and the source lasers. Without spectral tailoring, the pilot tone remains as double-sideband signal. Contrarily, oCS-SSB modulation pronounces the upper sideband of the pilot tone at +Ω relative to $f_v$ compared to its mirror frequency at the lower sideband, as presented in Fig. 7(b). The sideband suppression is 14 dB, which minimizes self-interference effects for the pilot reception and out-of-band crosstalk Ψ at the quantum plane. More importantly, the optical carrier at $f_v$ is suppressed by 23 dB so that detrimental in-band crosstalk Ξ at the quantum plane is greatly reduced at the receiver. Figure 7(b) also shows the effect of a transmitter-side polarization misalignment between I/Q modulator output and tributary-selective attenuator input, which leads to bleed-through of the data to the pilot plane.

The polarization-insensitive attenuator $A_T$ levels the compound PDM signal according to the desired launch $P_q$ of $\langle n \rangle$ = 4 photons/symbol for the quantum signal (i.e., its modulation variance $V_{mod} = 2 \langle n \rangle$). This leveling would also absorb the insertion loss of a high fan-out space switch that is required to time-share the CV-QKD hardware of the CO with many antenna sites.

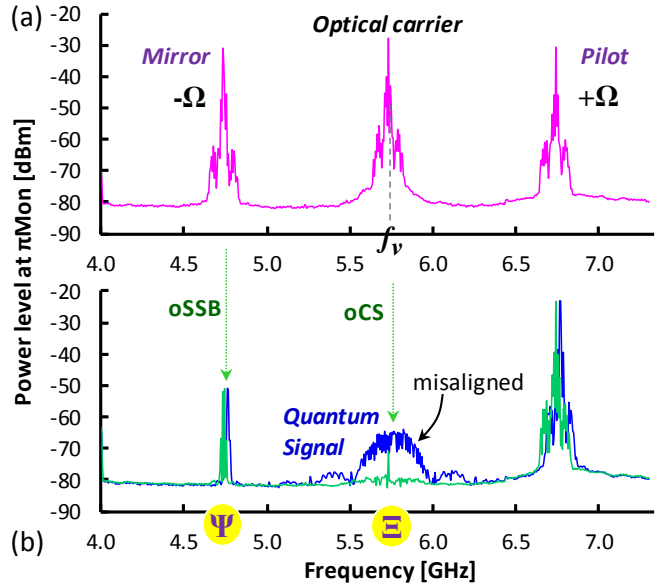

Fig. 7. Pilot tone spectrum for (a) double-sideband modulation and
(b) after sideband and carrier suppression for reduced crosstalk Ψ and Ξ during quantum signal reception.

In addition to the CV signal transmitter, the optical emitter is loading the transmission link with a comb of 11 classical channels ($\lambda_1 \ldots \lambda_{11}$). These channels reach from $\lambda_1 = 1548.51$ to $\lambda_{11} = 1560.61$ nm and feature an average power of 5.2 dBm/$\lambda$ at the transmission fiber input, as it is presented in the optical spectrum in Fig. 8. The chosen signal launch is motivated by the typical optical fronthaul budgets [47] and the reception sensitivities of the associated classical receivers [48]. This spectral setup allocates both, quantum and classical channels, in the C-band. A red/blue (R/B) filter for this waveband cleans the 1530-nm region before a 200-GHz optical add/drop (A/D) filter at 1531.9 nm, which combines the quantum and the classical signals at the output of the optical emitter. The WDM signal is then transmitted over an ITU-T G652.B-compatible fiber-based channel, as it is found in brown-field deployments. The transmission length of this standard single-mode fiber (SMF) has been chosen with 13.2 and 28.4 km, respectively. The kilometric loss of the fiber-based channel was 0.227 dB/km.

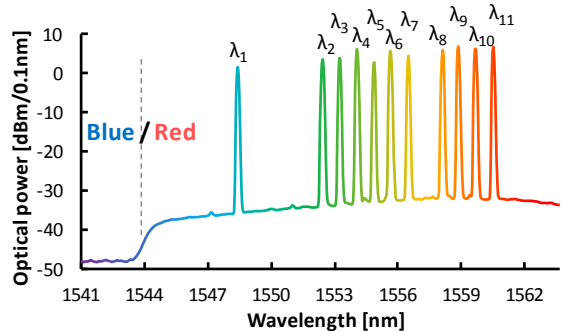

Fig. 8. Classical optical spectrum that is loading the mobile fronthaul link.

### D. Signal receiver

The CV receiver drops the classical signal through an A/D filter in order to avoid overloading of its detectors. A free-running and therefore optically unlocked laser with a linewidth of 10 kHz and a power of 16 dBm is employed as the LO. Together with the low linewidth of the laser at the optical transmitter, the ratio between laser linewidths to symbol rate $R_q$ is $<10^{-4}$ and therefore low enough to accomplish phase recovery at negligible penalties [49] during the subsequent signal processing step. The frequency mismatch $\Delta v$ between the source laser at the transmitter ($\lambda_T$) and the LO ($\lambda_R$) is minimized through current/temperature tuning, hence, coherent intradyne detection at $\Delta v \ll R_q$ applies. Continuous control of the input SOP to a PDM $\pi/4$-hybrid enables a crosstalk-free split of the two tributaries in the optical signal chain. In this way a quantum-heterodyne measurement is performed for the quantum signal by the hybrid, which directly yields the $I_{TE}$ and $Q_{TE}$ variables, pending frequency- and phase-offset correction. In parallel, the $I_{TM}$ and $Q_{TM}$ components of the pilot tone are received in the other PDM tributary.

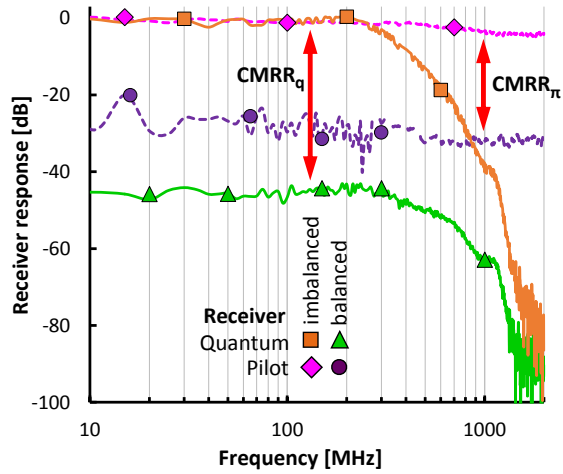

Fig. 9. Receiver response for the balanced homodyne detectors in quantum and pilot plane.

Balanced photodetectors are employed in both PDM planes. Their responsivities were 0.85 and 0.8 A/W, for the quantum-signal and pilot-tone receiver, respectively. Figure 9 presents the receiver response for the unbalanced (■,◆) and balanced (▲,●) case of common mode injection. The low-noise quantum receiver features a common-mode rejection ratio ($CMRR_q$) of 44.8 dB and a bandwidth of $B_{RX}$ = 315 MHz (■/▲), while the 10-GHz PIN/TIA pilot receiver has a $CMRR_\pi$ of 28 dB (◆/●). The bandwidth of the quantum receiver fits well to the desired roll-off sought for the response function $\rho(f)$, while it still enables monitoring of the pilot tone ($\Omega$) suppression for the purpose of continuous SOP control, which aims to maximize the signal at the quantum plane by minimizing the orthogonally polarized pilot tone.

A real-time oscilloscope retrieves the digitized representation of the electrical quantum and pilot tone signals. It has to be stressed that several clock sources for the waveform and pilot tone generator at the transmitter, and the analogue-to-digital conversion at the receiver, has been electrically synchronized. In a realistic deployment, the head- and tail-end CV-QKD equipment can build on the synchronization channels available through the wireless appliances.

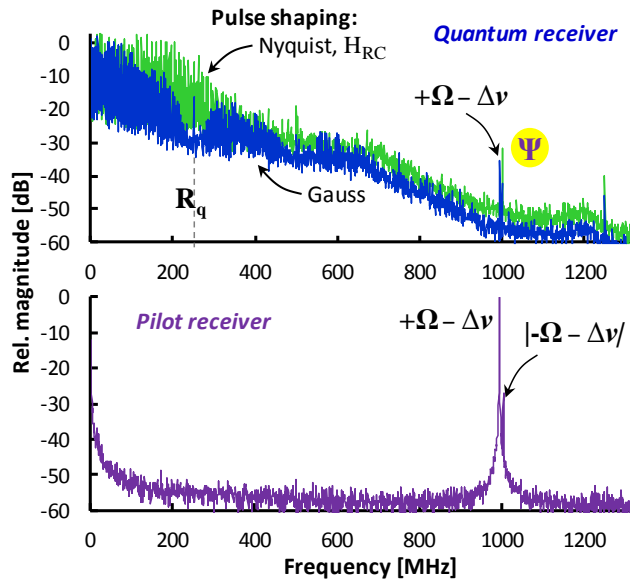

Fig. 10. Received uncarved quantum signal and pilot tone.

## E. Signal processing

Signal processing is performed off-line in the digital domain and includes signal conditioning, frequency/phase offset estimation and correction, and CV parameter estimation. The particular DSP blocks are explained in the following together with the resulting signals for the particular case of a high received power for quantum and pilot signal, in order to comprehensively visualize the DSP functions.

The received baseband quantum signal and the passband pilot tone are shown in Fig. 10. The spectra show the data modulation characteristics with Gaussian pulse shaping at the symbol rate $R_q$ = 250 MHz and the sharper roll-off for the Raised-Cosine pulse shaping at the doubled symbol rate $R_q$ = 500 MHz. The signal spectrum is determined by the Raised-Cosine filter function $H_{RC}$, as it is also shown in Fig. 5. Figure 10 also shows the single-sideband pilot tone with its residual mirror at $f = |-\Omega - \Delta v|$. The signals are first de-skewed with respect to each other and filtered by applying functions that have been matched to the symbol rate and the wavelength stability of the lasers, respectively. The frequency offset of the LO is then estimated from the actual and nominal pilot tone frequencies and is typically found below 10 MHz. After frequency offset correction, the signals are down-sampled from the acquisition rate of 20 GSa/s to $\Omega$ and $R_q$, respectively. This process takes advantage of the high SNR of the pilot tone to decide on the optimal decision sampling point. The carrier-phase $\varphi$ can then be estimated from the evolution of the unmodulated pilot tone in its I/Q plane. A typical phase evolution for signal transmission over 13.2 km is shown in Fig. 11 together with the pilot constellation diagram before (A) and after (B) phase compensation. A typical peak drift in phase experienced during the measurements was 1.72 rad/µs, which would correspond to 0.394 degrees/symbol for the lower symbol rate of $R_q$ = 250 MHz. The constellations for Gaussian (A-D) and Nyquist (E-H) pulse shaping shown in Fig. 12(a) have been acquired for an excessive level of 125 photons/symbol delivered to the receiver, in order to demonstrate the pilot concept. The information on the phase evolution, which is considered as reliable due to the high SNR, is applied for correcting the phase noise of the uncompensated quantum signal (C,G). This step yields the four phase-corrected in the quantum I/Q plane (D,H). Figure 12(b) presents the same diagrams for the desired modulation variance of 4 photons/symbol at the transmitter output.

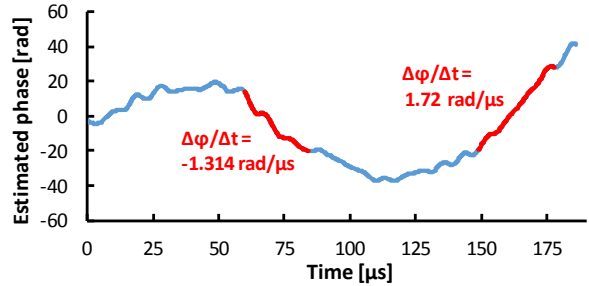

Fig. 11. Typical optical phase evolution after transmission over 13.2 km and coherent detection at the CV receiver.

Figure 13(a) shows a demodulated four-state, 500-MHz Nyquist-shaped quantum signal. In order to obtain a measure for the excess noise $\xi$ and also the channel transmission $T$, a parameter estimation is performed on the recovered quantum data. With this, the Holevo information can be estimated, which gives an upper bound of the mutual information between a potential eavesdropper and the receiving party.

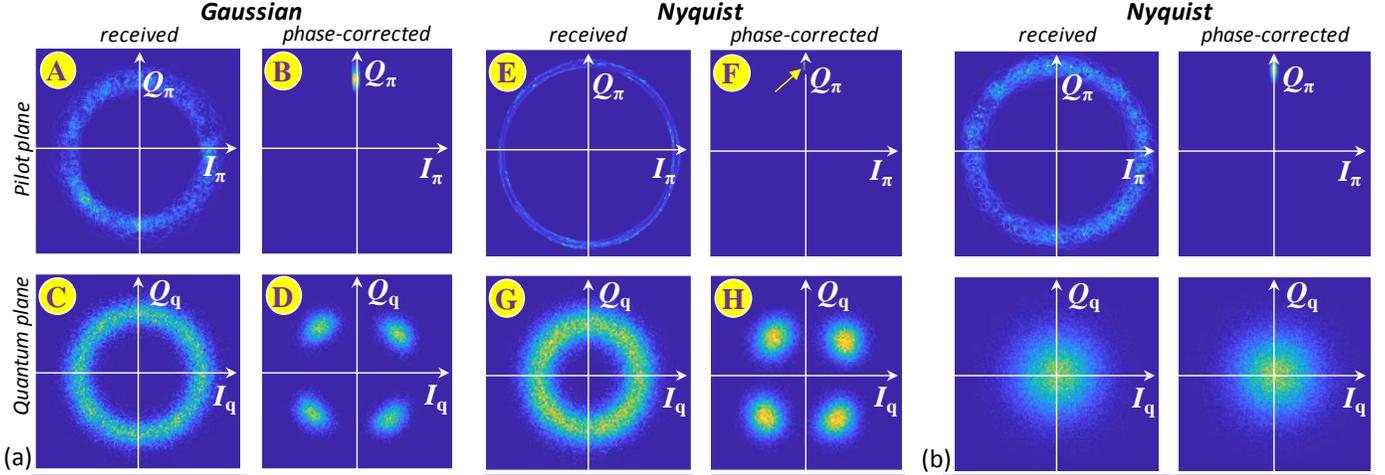

Fig. 12. Constellation diagrams before and after phase recovery in the pilot and quantum signal planes. Results are shown for (a) Gaussian (A-D) and Nyquist (E-H) pulse shaping and an excessive signal launch of 125 photons/symbol, and (b) for low modulation variance at the transmitter. For the last, the phase recovery of the quantum signal, though effective, is hard to recognize from the constellation diagram due to the low power.

## V. Results and Discussion

### A. Excess Noise

We use the excess-noise parameter as a primary performance indicator since any CV-QKD realization, regardless which exact protocol is chosen, is sensitive to noise. The excess noise is the quadrature variance in addition to the obligatory quantum shot noise with reference to the receiver input. The parameter estimation is based on [45]. It computes the conditional variance (i.e. the variance of Bob's samples conditioned on Alice's samples) and subtracts the quantum shot noise from it in order to yield the excess noise. We take $2^{17}$ data symbols into account for each of the measurements. A set of calibration measurements was done for the shot noise and the electronic TIA noise to yield the excess noise. Two cases are considered: First, the total excess noise $\xi$ and, second, the excess noise $\xi_T$ excluding the TIA noise in a scenario where the receiver is placed at a trusted location. This latter case builds on the relaxed security assumption that the receiver is considered as a trusted device which does not contribute to the information obtained by a potential eavesdropper [50].

Table III lists the obtained excess noise values $\xi$ and $\xi_T$ in quantum shot-noise units (SNU) inherent to the LO. Results are shown for a link reach of 13.2 km, as it is common for urban and mobile fronthaul scenarios. For Gaussian pulse shaping at $R_q$ = 250 MHz the total excess noise amounts to $\xi$ = 0.0146 SNU and 0.0145 SNU for uncarved and carved signal transmission, respectively. When loading the link with 11 classical channels, the excess noise increases slightly by 0.0024 SNU. This proves the feasibility for joint C-band transmission of both, quantum and classical signals. For the trusted receiver scenario the insecure noise reduces to $\xi_T$ = 0.0009 SNU and 0.0033 SNU for carved signal transmission without and with classical channels, respectively. This shows the strong impact of the TIA noise: the difference in excess noise, $\xi - \xi_T$, amounts to 1.36 %SNU when excluding the TIA noise from $\xi$, but the excess noise $\xi_T$ raises by only 0.24 %SNU when adding classical channels. Based on earlier findings [45], other sources of excess noise sources arising from sub-optimal common-mode rejection of direct-detection terms and relative intensity laser noise, analogue-to-digital quantization noise, or ripple in the electro-optic modulator response do certainly have an effect on the excess noise, however, they do so at a comparatively much smaller scale compared to receiver and Raman noise.

For a longer link reach of 28.4 km, the excess noise is $\xi$ = 0.0157 SNU without classical channels. It worsens to $\xi$ = 0.0232 SNU and $\xi_T$ = 0.0094 SNU under co-existence with the classical channels, meaning a 1.38× increase in total excess noise with respect to the shorter link of 13.2 km. With these, the longer fiber span causes a larger penalty when classical channels are present.

When Nyquist shaped transmission at $R_q$ = 500 MHz is applied, the excess noise over 13.2 km amounts to $\xi$ = 0.0172 SNU and 0.0212 SNU without and with classical channels. These values are slightly higher than found for Gaussian shaped transmission, however, the symbol rate is doubled.

Concerning other CV-QKD protocols, the phase-recovery based on the oCS-SSB pilot should work independently of the modulation alphabet used for the quantum signal. The difference would be found in the way the performance is evaluated. For example, in the case of a four-state protocol we can simply compare the variance of the four "blobs" with the shot-noise variance. For Gaussian modulation it would be required to compute the variance $var(x_A - x_B)$ of the differences between each pair of data points $x_A$ and $x_B$, see Eq. 8.25 in [45].

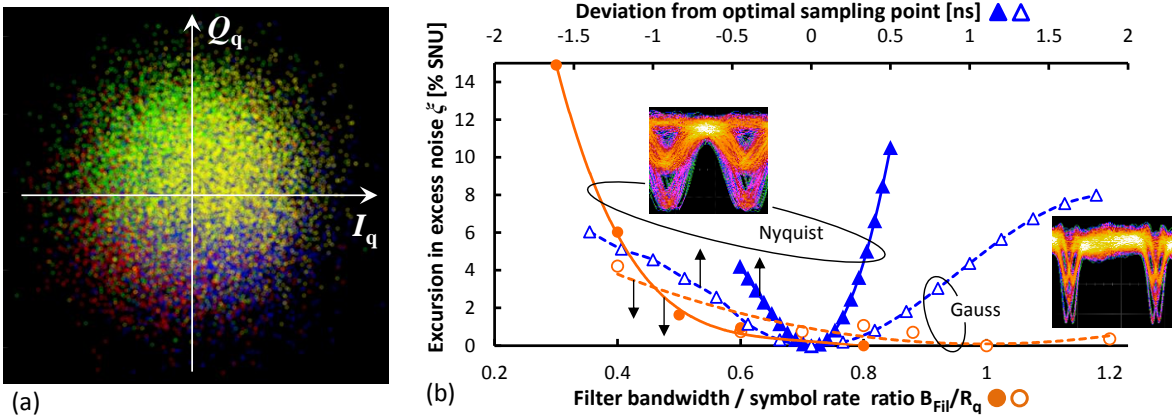

Fig. 13. (a) Demodulated Nyquist pulse-shaped 4-state quantum signal.
(b) Dependence of excess noise on receiver bandwidth and decision sampling offset.

## B. Filtering and Decision Sampling

The dependence of the reception performance on the reception bandwidth and the decision sampling accuracy is reported in Fig. 13(b) in terms of excess noise excursion. Results are provided for Gaussian pulse shaping at $R_q$ = 250 MHz (○,△) and Nyquist shaping at $R_q$ = 500 MHz (●,▲). It provides an indication on how the bandwidth limitation of the quantum receiver ($B_{RX}$ = 315 MHz) is imposing a limit for realizing high symbol rates $R_q$. For the presented investigation the reception bandwidth was determined by an additional low-pass filter ($B_{Fil}$) in the DSP chain. Although noise in the excess receiver bandwidth can be rejected through a spectrally narrow filter with $B_{Fil} < B_{RX}$, the received signal becomes quickly deteriorated when part of the signal spectrum is cut. As Fig. 13(b) shows, a penalty of 0.1% SNU in excess noise is experienced for 500-MHz Nyquist pulse shaping at a reception bandwidth of 380 MHz (●). For excessively large reception bandwidth $B_{Fil} > R_q$ and $B_{RX} > R_q$, excess noise leads to a slowly growing reception penalty. This is mainly observed for Gaussian pulse shaping (○) since for Nyquist pulse shaping at its higher symbol rate the incurred noise contribution from to the opto-electronic receiver is already limited due to the roll-off in its response (Fig. 9, ■) rather than the spectrally wide digital filter ($B_{Fil} > B_{RX}$).

The dependence of the penalty on an offset from the optimal decision sampling point is included in Fig. 13(b). An excess noise penalty of 0.1% SNU is incurred for an offset of 110 ps in case of Gaussian pulse shaping at 250 MHz (△). This underpins the importance of accurate decision sampling to suppress excess noise contribution due to DSP artifacts. The smaller Nyquist eye width, together with the higher symbol rate, require a very precise timing for symbol slicing (▲).

## C. Secure-key rate

The secure-key rate is derived off-line by means of simulations based on the experimentally measured values for excess noise, loss, modulation variance (i.e. signal amplitude) and symbol rate. The expected secure-key rate was estimated using the relation $K_S = R_q (\beta I_{AB} - \chi_{EB})$, where $\beta$ is the reconciliation efficiency, $I_{AB}$ is the mutual information between Alice and Bob, and $\chi_{EB}$ is the mutual information between Eve and Bob (often referred to as Holevo bound) [45]. To give such an estimation about the capability of the CV-QKD system to generate a secure key, further assumptions have been made. These include a Gaussian modulation alphabet, asymptotic keys, coherent attacks and reverse reconciliation with an efficiency of $\beta$ = 0.95 to obtain an asymptotic secure-key rate $K_S$ [45].

Under strict security assumption, meaning that the eavesdropper has control over the detector noise, a secure-key rate $K_S$ of 12 and 18.4 Mb/s can be obtained for a dark 13.2 km link employing Gaussian and Nyquist pulse shaping, respectively. These rates drop to 9.6 and 10.7 Mb/s when all classical channels are lit up. The obtained key rate would marginally exceed the demanded rate of 10 Mb/s, as initially outlined in Section II.B.

| Quantum signal Characteristic | $R_q$ = 250 MHz Gauss-shaped | | | $R_q$ = 500 MHz Nyquist-shaped | |
|---|---|---|---|---|---|
| Symbol carving | absent | present | | | |
| Classical channels, fiber illumination | 0 dark | 0 dark | 11 lit | 0 dark | 11 lit |
| SNR [1] | 0.38 | 0.28 | 0.12 | 0.16 | 0.17 |
| Total excess noise $\xi$ [% SNU] | 1.465 | 1.446 | 1.683 | 1.721 | 2.12 |
| Excess noise $\xi_T$ [% SNU] for trusted receiver | 0.111 | 0.092 | 0.329 | 0.115 | 0.514 |
| Key rate $K_S$ [Mb/s] | 11.8 | 12 | 9.59 | 18.4 | 10.7 |
| Key rate $K_T$ [Mb/s] for trusted receiver | 42.8 | 43.2 | 38.9 | 85.3 | 72 |

Table III. CV-QKD reception performance for fiber reach of 13.2 km.

However, under practically viable assumptions that the receiver is placed at a trusted location, a secret key can be established. A key rate $K_T$ of 43.2 Mb/s is obtained over 13.2 km of dark fiber under Gaussian shaped and pulse carved transmission at $R_q$ = 250 MHz. When the classical channels are present, the extra Raman noise contribution degrades the key rate to 38.9 Mb/s. The data rate can be improved through Nyquist-shaped transmission at $R_q$ = 500 MHz. The obtained key rates over dark and lit fiber are 85.3 and 72 Mb/s, respectively. These results underpin the feasibility of realizing QKD rates over mobile fronthaul distances that are clearly exceeding 10 Mb/s, using telecom-grade componentry.

For a longer reach of 28.4 km and assuming a trusted receiver, the increased loss of the dark fiber channel leads to a supported secure-key rate $K_T$ of 10.3 Mb/s under Gaussian-shaped transmission at $R_q$ = 250 MHz. However, the increased Raman noise lowers the secure-key rate to 2.76 Mb/s for a fully-loaded link. In this loss-limited regime the intrinsic loss contribution of a phase-diversity intradyne reception scheme is seen as the limiting factor to generate higher secure-key rates. Alternative solutions, such as coherent heterodyne detection with reduced intrinsic receiver loss, may prove more practical despite their spectral inefficiency.

*D. Modulated classical channels*

The classical channels had been left unmodulated for the presented transmission experiment. In order to estimate the impact of modulated classical channels, the crosstalk arising from a modulated WDM comb has been analyzed. To do so, the out-of-band WDM channels had been modulated with polarization-multiplexed quadrature phase shift keyed (PM-QPSK) data signals at a 10 Gbaud symbol rate using pseudo-random bit sequences. The classical signals have been injected into the receiver without additional quantum channel, at an average optical power of -15.4 dBm/λ and thus at a typical margin-adjusted reception sensitivity [48]. The LO has been left at its spectral displacement from the classical WDM channels, according to the reception of the quantum signal.

Under this condition there is no spectral component of the classical signal falling within the reception band of the receiver, which is defined by the LO. However, in case of imperfect detector balancing, common mode noise will arise due to direct detection terms [51]. We have therefore investigated the dependence of crosstalk as a function of the balancing quality, which can be expressed through the CMRR [52]. However, the high data rate of the classical channels would lead to just a few crosstalk harmonics within the much smaller quantum receiver bandwidth. For this reason, this investigation was conducted using the pilot receiver, which shows a similar bandwidth as the classical signal spectra.

Figure 14 presents the received RF spectrum for a partially balanced pilot receiver with a certain $CMRR_\pi$. The received signal for the balanced receiver with a high $CMRR_\pi$ of 28.3 dB remains without spectral modulation components due to the out-of-band classical channels. When degrading the CMRR further, first spectral components appear in the spectrum (Fig. 14(b)). In the worst case of a rather unbalanced detector with a low $CMRR_\pi$ of 9.5 dB, data harmonics of the classical channels are clearly dominating the signal spectrum. This shows that out-of-band WDM channels can contribute to the excess noise. However, considering that the quantum receiver is operated at a much higher $CMRR_q$ of 44.8 dB when balanced, we do not expect an excess noise contribution due to modulated classical channels.

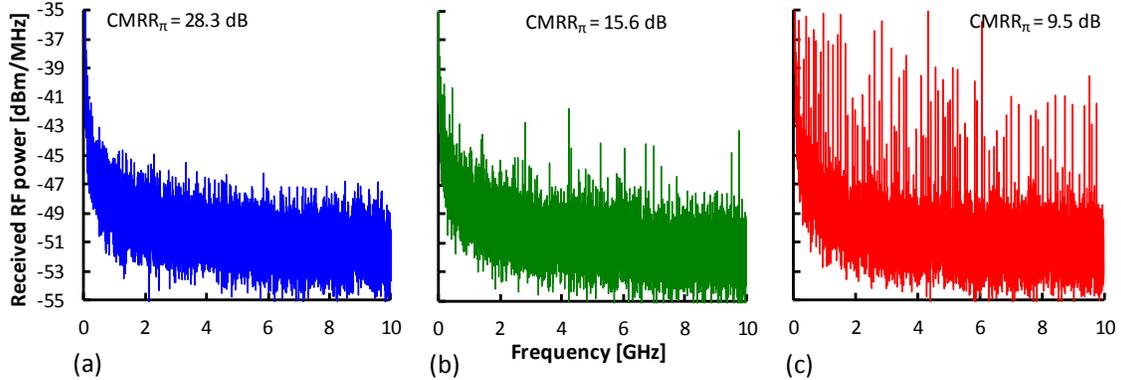
Fig. 14. Received signal spectra of the broadband pilot tone receiver under partial balancing. Results are shown without CV transmission but in presence of spectrally remote load channels.

## VI. Conclusion

We have experimentally demonstrated a coherent transmission system for a high-rate CV-QKD scheme with free-running LO at the receiver. Accurate frequency estimation and optical phase estimation has been accomplished through use of a pilot tone that is multiplexed in both, frequency and polarization. Crosstalk due to the optical carrier component of this training signal has been mitigated through spectral tailoring based on optical carrier suppression, while self-interference due to coherent intradyne reception has been suppressed through single-sideband modulation of the pilot tone. The use of such a scheme allows for high symbol rates of 250 MHz for the quantum signal, which, despite the limited quantum receiver bandwidth, can be further extended to 500 MHz by means of Nyquist pulse shaping. The transmission performance has been assessed over a fiber-based fronthaul link with a reach of 13.2 km. Low excess noise allows for the generation of secure keys in a trusted receiver scenario. The obtained asymptotic secure-key rate is 85.3 Mb/s over the dark fronthaul. In co-existence with 11 carrier-grade classical channels that are spectrally allocated in the same waveband as the quantum signal, being spectrally spaced by only 20 nm, a secure key can be still established at 72 Mb/s. Though already complying with the requirements of mobile fronthaul links, a reduction of intrinsic receiver losses is paramount for further scaling of the link reach or loss.

In case of a 5G deployment scenario where either Alice or Bob is shared at the CO among many antenna sites, the achieved key rate is high enough to accommodate a periodic AES key exchange. Since signal processing is hosted at the CO, it is likewise to locate Bob as the CV receiver centrally, while many Alices as the CV transmitters are being field installed. Such a deployment strategy would also overlap with a migration towards coherent PON technology, which builds on similar transmitter architectures. Towards this direction, multi-purpose photonics is seen as a shuttle to incorporate QKD functions in classical transceivers.

## VII. Acknowledgement

This work was supported in part by funding from the European Union's Horizon 2020 research and innovation programme under grant agreement No 820466 and the European Research Council (ERC) under grant agreement No 804769.